\documentclass[aps,pra,amsmath,amssymb,11pt,superscriptaddress]{revtex4-2}
\usepackage{amsthm}  
\usepackage{amsmath,amssymb,amsfonts,graphicx,bbm,mathtools,braket,color,dsfont}
\usepackage{lipsum}
\usepackage{xcolor}
\usepackage{soul}

\begin{document}

\title{Link prediction with swarms of chiral quantum walks}

\author{Gaia Forghieri}\email{gaia.forghieri@unimi.it}
\affiliation{Dipartimento di Fisica, Università di Milano, I-20133 Milan, Italy}

\author{Viacheslav Dubovitskii}
\affiliation{Algorithmiq Ltd., Kanavakatu 3 C, FI-00160 Helsinki, Finland}

\author{Matteo A. C. Rossi}
\affiliation{Algorithmiq Ltd., Kanavakatu 3 C, FI-00160 Helsinki, Finland}

\author{Matteo G. A. Paris}
\affiliation{Dipartimento di Fisica, Università di Milano, I-20133 Milan, Italy}

\date{\today}

\begin{abstract}
Reconstructing protein-protein interaction networks is a central challenge in network medicine, often addressed using link prediction algorithms. Recent studies suggest that quantum walk-based approaches hold promise for this task. In this paper, we build on these algorithms by introducing chirality through the addition of random phases in the Hamiltonian generators. The resulting additional degrees of freedom enable a more diverse exploration of the network, which we exploit by employing a swarm of chiral quantum walks. Thus, we enhance the predictive power of quantum walks on complex networks. Indeed, compared to a non-chiral algorithm, the chiral version exhibits greater robustness, making its performance less dependent on the optimal evolution time—a critical hyperparameter of the non-chiral model. This improvement arises from complementary dynamics introduced by chirality within the swarm. By analyzing multiple phase-sampling strategies, we identify configurations that achieve a practical trade-off: retaining the high predictive accuracy of the non-chiral algorithm at its optimal time while gaining the robustness typical of chirality. Our findings highlight the versatility of chiral quantum walks and their potential to outperform both classical and non-chiral quantum methods in realistic scenarios, including comparisons between successive versions of evolving databases.
\end{abstract}

\maketitle

\section{Introduction}

Link prediction is a fundamental problem in network science, with applications on network reconstruction spanning from friendship recommendations in social media \cite{adamic2003,daud2020,kumari2022} to biological systems \cite{kovacs2019,abbas2021,yuen2023,almusawi2025}. In regards to the latter, protein-protein interaction (PPI) networks, also called \textit{interactomes}, constitute a challenge of critical importance. Indeed, these networks map all known interactions between proteins occurring within the cell of an organism \cite{goel2012,das2012,alonsolopez2019,luck2020,oughtred2021}. Therefore, they are crucial in understanding cellular processes and disease mechanisms \cite{richards2021,wang2022,kumar2024}, ultimately enabling optimal drug design \cite{loscalzo2023,liu2024,campsfajol2025}. However, the currently available data on experimentally validated interactions is still largely incomplete, with human interactome databases estimated to cover only a small fraction of the whole structure ($\lesssim 10\%$) \cite{luck2020,dimitrakopoulos2022,kosoglu2024}.

Due to the great sparsity and incompleteness of PPI networks, a great deal of link prediction methods have been tested on them \cite{kovacs2019,nasiri2021,jha2022,hu2024}. In this context, accurate link prediction is essential, since it provides a basis for guiding experimental validation with an efficient selection of candidate links that are most likely to exist. Among the possible methods, those based on continuous-time random walks (RW's) have proven to be particularly effective \cite{liu2010,curado2020,xia2020,vital2024}. Indeed, the evolution of a walker along the network explores its whole topology, potentially predicting links between far-distant nodes.

Recent works have extended the RW-based method to its quantum counterpart, introducing continuous-time quantum walks (QW's) \cite{goldsmith2023,moutinho2023,moutinho2024}. Thanks to its quantum nature, the QW allows for phenomena that are classically absent, most notably superposition and interference between different trajectories of the walker. Consequently, the evolution of a QW is drastically different from that of the RW and, under the appropriate conditions, can be faster and even reach the ballistic regime \cite{aharonov1993,mulken2011,wang2020,kadian2021,qiang2024}. This has proven to greatly improve the efficiency in the exploration of the network topology, consequently enhancing the performance for link prediction. Additionally, the same formalism can be adopted for many other problems in the scope of network science, such as disease gene prioritization \cite{saarinen2024,dubovitskii2025}, and more.

That said, QW-based link prediction methods up to now have relied on symmetric, real-valued Hamiltonians directly derived from the adjacency or Laplacian matrices of the graph. However, this is an unnecessary restriction. Instead, a more general Hamiltonian generator can be built through the introduction of complex phases for each edge of the network. {\color{black}From a physical perspective, the complex phases can be interpreted as effective synthetic gauge fields acting on the graph \cite{boada2017,cedzich2019,razzoli2019}, modifying the relative phases accumulated along closed paths and thus altering the constructive or destructive interference between different trajectories.} In this context, we talk about \textit{chiral} QW's \cite{lu2016,frigerio2021,frigerio2022}, since these phases induce a directional bias in the evolution of the {\color{black}and explicitly break time-reversal symmetry. As a consequence, chirality gives access to a qualitatively different interference structure with respect to standard non-chiral QW’s, rather than acting as a mere perturbation of the underlying dynamics. As a result, chiral quantum walks can probe hidden paths and long-range correlations in the graph that are not accessible to quantum walks generated by standard symmetric Hamiltonians}. This property can dramatically alter transport properties on graphs, and is consequently a key ingredient in enhancing QW-based algorithms, such as quantum transport \cite{frigerio2022,apers2022,cavazzoni2022,frigerio2023,annoni2024,finocchiaro2025}, state transfer \cite{khalique2021,bottarelli2023,ragazzi2025,cavazzoni2025}, and so on.

In this paper, we study the effects of chirality on the link prediction problem, by exploiting the wide range of dynamics offered by the phases as new degrees of freedom. Specifically, we combine multiple QW's with different directional biases, to form what we call a \textit{swarm} of chiral QW's. In this way, we aim at enhancing the exploration of sparse and irregular networks {\color{black}by effectively sampling a broader space of admissible quantum dynamics. Consequently, this will allow to complement} the non-chiral evolution {\color{black}in the exploration of possible links to predict, providing a contribute that exceeds that of perturbative methods}. We compare the chiral and non-chiral performances on multiple real-world PPI networks, and demonstrate in which regimes the use of chirality allows to outperform the non-chiral method. 

Our results demonstrate that chiral quantum walks may possess the potential to enhance link prediction performance, particularly in largely incomplete PPI networks. By combining multiple chiral walks into a swarm, we introduce complementary dynamics that bring an improved stability of the performance over time and, under certain conditions, even outperform the non-chiral performance at the optimal time. 
These findings suggest that chirality is a powerful tool for overcoming the limitations of conventional quantum walks in reconstructing interactomes, often complementing other best known methods from literature. {\color{black}Importantly, this increased robustness makes the chiral method particularly suitable for realistic link prediction scenarios, such as PPI network reconstruction, where the optimal evolution time cannot be determined \textit{a priori}.}

The paper is organized as follows: Section \ref{sec:methods} introduces the theoretical framework for chiral quantum walks and the metrics used for link prediction. Section \ref{sec:results} presents our results, beginning with a preliminary analysis on the behavior of swarms of chiral QW's in comparison to a single non-chiral QW, both at reference unitary time and in terms of dynamics. We then compare our method to other approaches from literature and analyze its efficacy across diverse PPI networks. At the end, we briefly focus on link prediction applied on the comparison of different temporal versions of the same network, to better assess the performance on real-case scenarios. Finally, Section \ref{outro} discusses implications and future directions, while Appendices provide additional technical details on dynamical distances between walker evolutions and performance metrics.

\section{Methods}\label{sec:methods}
We consider complex networks modeled by unweighted and undirected graphs $G(V,E)$, with $V$ the set of nodes of size $n$ and $E$ the set of edges. 
The adjacency matrix of $G$ is defined as an $n\times n$ matrix:
\begin{align}
    A=(A_{jk})=\begin{cases}
        1 \quad &{\rm if}\; (j,k)\in E\,, \\
        0 \quad &{\rm if}\; (j,k)\notin E\, .
    \end{cases}
\end{align}
Another important definition is the degree of node $j$, $d_j=\sum_kA_{jk}$, defined as the number of links emanating from node $j$.


\subsection{Continuous-time quantum walks}

The description of the QW requires the definition of an $n$-dimensional Hilbert space $\mathcal{H}$. Each vector of its orthonormal basis $\{\left|j\right\rangle\}_{j=1}^n$ represents a state completely localized on node $i$ of $G$. Within this formalism, the occupation probability of each node at time $t$ is $p_j(t)=|\left\langle j|\psi(t)\right\rangle|^2$, where $\left|\psi(t)\right \rangle$ is a generic quantum state in $\mathcal{H}$. The QW evolution is described by the Schr\"odinger equation:
\begin{align}
    \left|\psi(t)\right\rangle = \mathcal{U}(t)\left|\psi(0)\right\rangle\, ,
\end{align}
where $\left|\psi(0)\right\rangle$ is the initial state of the system, and $\mathcal{U}(t)=e^{-iHt}$ is the unitary evolution operator (fixing $\hbar=1$). Thus, the transition probability from nodes $j$ to $k$ reads $P_{jk}(t)=\left|\left\langle j|\, \mathcal{U}(t)|k\right\rangle \right|^2$. 

The only conditions in choosing the Hamiltonian generator $H$ for the evolution of the system are for it to be Hermitian, so that $\mathcal{U}$ is unitary, and compatible with the network's topology. Consequently, a possible choice is to directly associate $H\equiv A$. 
However, one can generalize the concept to include complex phases on the off-diagonal elements of $A$:
\begin{align}\label{eq:chiral_ham}
    H_{\phi}=(H_{\phi,jk})=\begin{cases}
        e^{i\phi_{jk}} \quad &{\rm if}\; (j,k)\in E\,, \\
        0 \quad &{\rm if}\; (j,k)\notin E\, ,
    \end{cases}
\end{align}
with $\phi_{kj}=-\phi_{jk}$. Since these introduce a directional bias in the dynamics of the walker {\color{black}by breaking time-reversal symmetry}, we talk in this case of a \textit{chiral} quantum walk (c-QW) \cite{lu2016}. The special case in which all phases are zero now takes the name of \textit{non-chiral} quantum walk (nc-QW). One can immediately notice how the introduction of such phases implies the existence of infinitely many c-QW's for each nc-QW. {\color{black}This means that, even for a fixed underlying graph, one can access a large family of inequivalent quantum dynamics, all compatible with Hermiticity and network topology.} This opens up the possibility to optimize the quantum advantage for the task of interest \cite{frigerio2022}, {\color{black}and, in the context of link prediction, enlarges the set of possible interference patterns through which the walker probes the graph structure.}

\subsection{Metrics for link prediction}
The objective of link prediction is to infer the missing links in $G$ based on its known structure. Therefore, a link prediction algorithm usually consists of a scoring scheme that produces a ranking of all possible pairs of non-connected nodes in $G$. Ideally, the higher the score of the pair is, the higher the probability that a link is present in the unknown structure of $G$. Link prediction methods that involve the evolution of a nc-QW on complex networks have already been studied \cite{liang2022,goldsmith2023,moutinho2023,moutinho2024}, and typically evaluate the score of each non-connected pair $(j,k)$ as a function of the transition probability between nodes $j$ and $k$ after an evolution of time $t$. Indeed, these methods rely on the intuition that a link is more likely to occur between two nodes if the walker is more likely to move from one to the other. We will refer to the scoring system defined previously by Goldsmith et al. \cite{goldsmith2023}:
\begin{align}\label{eq:score1}
    S(j,k;t) = P_{jk}(t)(d_j+d_k)\, ,
\end{align}
for $j\neq k$, which also takes into account the degree of the nodes involved, as high-degree nodes are the most likely to be connected to other nodes. {\color{black}Notice that the evolution time $t$ plays the role of a hyperparameter in QW-based link prediction, as it controls the extent of the exploration. Short times correspond to a transient, largely local regime, while long times are dominated by recurrent interference effects involving the full spectrum of the Hamiltonian. As discussed later in Sec. \ref{sec:dynamic}, physically relevant timescales are therefore expected to be of the order of the inverse spectral gap of the Hamiltonian.}

\begin{figure}
    \centering
    \includegraphics[width=\linewidth]{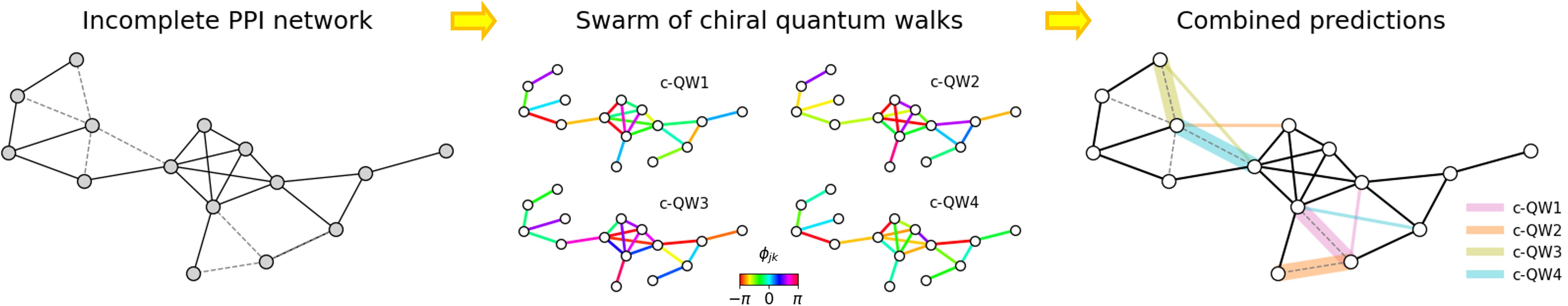}
    \caption{\color{black}Schematical representation of the link prediction process with a swarm of c-QW's. Starting with a PPI network with a number of missing links (left panel), we build several Hamiltonian generators, each corresponding to a c-QW, by assigning random phases to each known edge (central panel). Each walker assigns a score to every candidate link through its evolution. Final predictions are obtained by combining these outcomes and associating to each potential edge the maximum score achieved across the swarm (right panel). The colored edges in the right panel represent predicted links above a certain score threshold. Their width is proportional to the final score, and the color represents which walker corresponds to the considered maximum score.}
    \label{graphical_abstract}
\end{figure}

As we can consider multiple QW's evolving on $G$, we can imagine that each will explore the network differently, and potentially predict missing links in areas that are unexplored by the others. This is a direct consequence of the directionality imposed on the walker dynamics by the chiral phases in the Hamiltonian from Eq. \eqref{eq:chiral_ham}. Because of this, a possible strategy to try and maximize the performance of link prediction is to exploit a \textit{swarm} of quantum walks, and keep for each non-connected pair the maximum score predicted by each QW. We can thus define, for $M$ chiral walkers, an overall chiral score:
\begin{align}\label{eq:score2}
    S_{c}^{M}(j,k;t)=\max(\{S_{m}(j,k;t)\}_{m=1}^{M})\, ,
\end{align}
where $S_{m}(j,k;t)$ is the score obtained by the $m$-th c-QW. {\color{black}In this way, we can combine the best prediction from each walker, as schematically illustrated in Fig. \ref{graphical_abstract}.} In the same way, one can combine results from both the non-chiral and chiral cases, obtaining a total score of:
\begin{align}\label{eq:score3}
    S_{tot}^M(j,k;t)=\max(S_{nc}(j,k;t),S_c^M(j,k;t))\,
    ,
\end{align}
where $S_{nc}(j,k;t)$ is the score from the nc-QW.

The whole structure of PPI networks is generally unknown. Therefore, we test the performance of our link prediction algorithm on a certain dataset by assessing its capability to retrieve a fraction of randomly removed links using the remaining information on the dataset. That is to say, for each considered network we randomly remove $N\%$ of the existing links. All removed links are considered positive test cases, while the non-edges from the original dataset are negative test cases. Subsequently, we run our prediction algorithm on the remaining structure of the graph. All the non-edges are ranked in descending order according to the scores from Eqs. (\ref{eq:score1}-\ref{eq:score3}): the higher the score, the more likely is the link to exist. Finally, the edges whose score is higher than a certain cut-off threshold are considered positive predictions, while all the other edges are negative predictions.

This approach allows us to define the following metrics for each value of the threshold:
\begin{align}
    \rm precision= \frac{TP}{TP+FP}\, ; \quad recall = TPR = \frac{TP}{TP+FN}\, ; \quad FPR = \frac{FP}{FP+TN}\, ;
\end{align}
where TP = true positives, FP = false positives, TN = true negatives, FN = false negatives. TPR/FPR stand for true/false positive rate, respectively. The figures of merit we will refer to in Sec. \ref{sec:results} are the area under the receiver-operator characteristic curve (AuROC, TPR vs. FPR) and the area under the precision-recall curve (AuPR, precision vs. recall). Both of these metrics have been extensively used in link prediction and more general binary classification problems. Of the two, the AuROC is the more extensively used in literature, thanks to its easier interpretability \cite{zhou2021}. Indeed, it can be interpreted as the probability of correctly predicting a link chosen at random from the positive set, over a wrong link chosen at random from the negative set. On the other hand, the AuPR focuses only on the positive class, and is consequently the more relevant for scenarios where the ratio between positive and negative cases is extremely small. This is the case for PPI networks, which generally present high sparsity. Additionally, it is the more suited for evaluating early retrieval performance, an extremely important aspect for biological applications \cite{saito2015}.

\section{Results}\label{sec:results}
\begin{figure}[h!]
    \centering
    \includegraphics[width=0.5\linewidth]{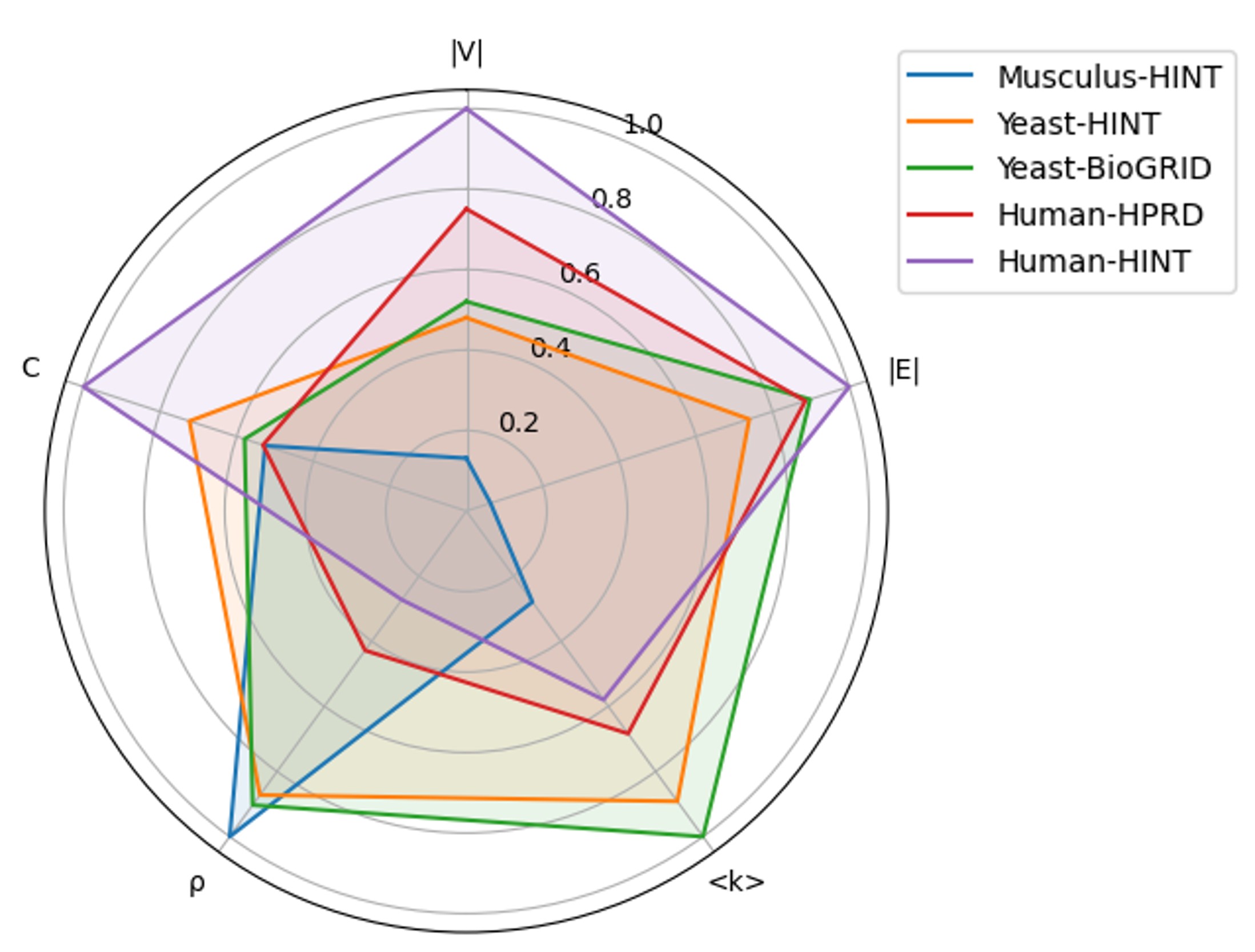}
    \caption{Radar chart of normalized metrics for each network from Table \ref{tab:param}. Each quantity $q\in\{|V|,|E|,\langle k\rangle,\rho,C,A\}$ is normalized through the equation $q^{\rm Norm}=q/q^{\rm max}$, where the maximum is evaluated among all considered networks.}
    \label{fig:graphs}
\end{figure}

In our study, we considered five different networks. Three of them are from the HINT database \cite{das2012}, version 2024-6: Musculus-HINT, corresponding to the common house mouse; Yeast-HINT, i.e., the \textit{Saccharomyces Cerevisiae} yeast commonly used for baking; and Human-HINT. We also use the Yeast-BioGRID network, still for S. Cerevisiae, from the BioGRID database \cite{stark2006}, version 4.4.248, compiled on July 25th, 2025, to compare results on the same organism from different databases. For the same reason, we also include the Human Protein Reference Database (HPRD) from its latest update \cite{keshava2009}. For each of these networks, we consider only the largest component. Additionally, we don't consider self-edges, since these have little to no influence on the performance \cite{goldsmith2023}, and are not modified by the introduction of chiral phases. Some statistics of the resulting networks are reported in Table \ref{tab:param}, and represented graphically in Fig. \ref{fig:graphs}.

\begin{table}[h!]
    \begin{tabular}{|l|c|c|c|c|c|c|}
\hline
    Network & $|V|$ & $|E|$ & $\langle k\rangle$ & $\rho$ & $C$\\
\hline
    Musculus - HINT & 1,486 & 2,423  & 3.261 & 0.0022 & 0.108 
    \\
    Yeast - HINT    & 5,438 & 28,319 & 10.415 & 0.0019 & 0.148 
    \\
    Yeast - BioGRID & 5,892 & 34,418 & 11.683 & 0.0020 & 0.119 
    \\
    Human - HPRD    & 8,498 & 33,935 & 7.987 & 0.0009 & 0.109 
    \\
    Human - HINT    & 11,319 & 38,332 & 6.773 & 0.0006 & 0.205 
    \\
\hline
    \end{tabular}
    \caption{Some properties of the networks that were tested. $|V|$ : number of nodes, $|E|$ : number of edges (excluding self-edges), $\langle k \rangle$ : average degree, $\rho$ : network density, $C$ : average clustering coefficient.}
    \label{tab:param}
\end{table}

In the following, we study the $k$-fold cross-validation for $10\%$ ($k=10$), $20\%$ ($k=5$) and $50\%$ ($k=2$) link removal. We repeat the cross-validation $r=2$ times for $20\%$ and $r=5$ times for $20\%$, in order to average the results on a total of $10$ trials for each case. Firstly, in Sec. \ref{sec:musculus}, we compare the performances of nc-QW's and c-QW's on Musculus-HINT, the smallest network considered in this work (see Table \ref{tab:param}), to get a general idea on how to efficiently build a link prediction algorithm based on chirality. Here, in Sec. \ref{sec:static} we first focus on how to better exploit multiple complemental c-QW's at a reference unitary time. We then analyse the dynamical behaviour of the performance in Sec. \ref{sec:dynamic}. Finally, in Sec. \ref{sec:benchmark} we study the performance on multiple PPI networks, comparing our QW-based algorithms with other well-known approaches from literature, with an additional focus on version comparison in Sec. \ref{sec:version}.

\subsection{Swarm performance}
\label{sec:musculus}

\subsubsection{Making the most of chiral complementarity}\label{sec:static}

We start our analysis by considering the effects of the contribution of multiple c-QW's to our algorithm, which we refer to as a \textit{swarm} of chiral walkers. We consider $t$ as a hyperparameter of our problem, on which the performance can be optimized. However, understanding the optimal value of $t$ requires \textit{a priori} knowledge on the evolution of the performance, something which in general varies wildly depending on the graph structure and is not usually known in real-life scenarios. {\color{black}The choice $t=1$ therefore represents a realistic one‑shot reference. We will show later that this time is generally close to the optimal time interval, so that the chiral swarm retains near‑optimal performance} (see the later analysis in Sec. \ref{sec:dynamic}). 

\begin{figure}[h!]
\centering
    \includegraphics[width=0.7\linewidth]{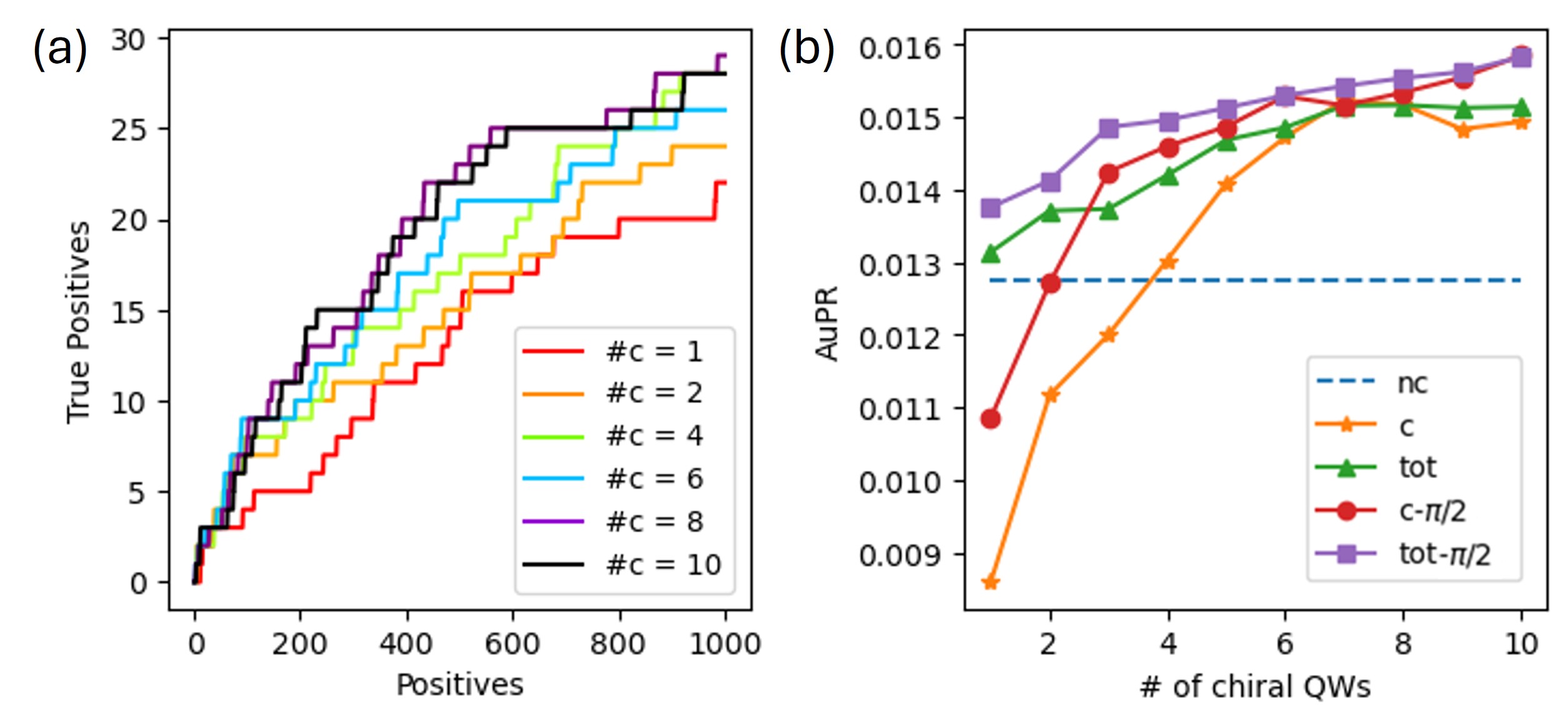}
    \caption{Performance of link prediction at $10\%$ link removal on the Musculus-HINT network as a function of the number of considered chiral QW's, at fixed $t=1$. Panel (a) shows the number of links correctly predicted (true positives) among the first predicted non-edges with highest score (positives) for a single fold, considering an increasing number of c-QW's with random phases and without the nc-QW's contribution. Panel (b) shows the convergence of the average AuPR from 10-fold cross-validation. The label `nc' stands for non-chiral; `c' and `tot' refer to the chiral case with random phases, respectively with and without the contribution from the nc-QW; the suffix `$\pi/2$' is used to differentiate the case with phases sampled as random multiples of $\pi/2$ rather than from the continuous interval $[-\pi,\pi]$ (with no suffix).}
    \label{fig:conv}
\end{figure}

In Fig. \ref{fig:conv}(a) we show the number of true positive predictions at each threshold with only chiral contributions to the score [see Eq. \eqref{eq:score2}]. These trials were performed at $10\%$ link removal for the smallest network we considered in our study, i.e. Musculus-HINT (see Table \ref{tab:param}). We considered an increasing number of c-QW's with phases sampled randomly in the interval $\phi_{jk}\in[-\pi,\pi]$. These results show that the early retrieval rate of the link prediction algorithm steadily increases with the number of considered c-QW's. This behaviour can be explained by the possibility of exploring a wider range of dynamics thanks to the directionality introduced by the chiral phases. Indeed, due to the highly complex structure of the considered network, the evolution of a single walker tends to concentrate on specific areas of the graph, and consequently miss the presence of potential links in other regions. Instead, a second walker with different chirality will potentially explore the graph differently, and consequently recover information that the first walker missed. Eventually, a swarm of a sufficient number of walkers is likely to explore the complete structure of the graph, thus saturating the performance, as seen in Fig. \ref{fig:conv}(a) for $\sim 10$ c-QW's.

We then show in Fig. \ref{fig:conv}(b) the average AuPR from the 10-fold cross-validation on the same network. In this picture, we show results obtained from evaluating the score either with chiral contributions alone [Eq. \eqref{eq:score2}], or by including the non-chiral case as well [Eq. \eqref{eq:score3}]. The reported curves confirm our previous considerations. Indeed, in both cases the AuPR increases as a function of the number of c-QW's making up the swarm, until it mostly converges for $\sim 10$ c-QW's. Most importantly, as the number of walkers increases, the performance eventually exceeds that of the nc-QW alone, while a single c-QW on average behaves worse than its non-chiral equivalent. This implies that the power of random chirality stands in the possibility of combining multiple dynamics, while the nc-QW is generally the best choice when using a single walker on its own. Additionally, we show that the inclusion of the nc-QW significantly affects the performance by increasing the AuPR at each step. This confirms that the non-chiral evolution has a particular importance in the exploration of the graph. Indeed, the absence of directionality in the Hamiltonian generator makes it so that the nc-QW already explores in a relatively uniform manner all regions of the graph. The same observations are also valid for the AuROC, though to a lesser scale (see Appendix \ref{sec:auroc}).

After these considerations, we introduced a second sampling method for the phases in Eq. \eqref{eq:chiral_ham}. This consists in only sampling multiples of $\pi/2$, and rearranging them so that the sum of the phases on each row of $H_{\phi}$ is as close to 0 as possible. This choice ensures a set of c-QW's making up the swarm with evolutions that are, on average, as different as possible from the nc-QW and from one another (see Appendix \ref{sec:distance}). In other words, this second sampling method (suffix `$\pi/2$') ensures a more efficient exploration of the configuration space made available by chirality. Consequently, as the AuPR shows, this method consistently enhances the performance, both with and without the non-chiral contribution. 

Overall, the results from Fig. \ref{fig:conv} show that the nc-QW is the one that, on its own, best explores the graph for link prediction purposes. However, a swarm of c-QW's can match, and eventually improve, the performance of the nc-QW alone. Specifically, chiral swarms that introduce the largest variety with respect to the non-chiral evolution bring the best performance overall, at least when considering $t=1$ as a reference time. The next question is whether or not this behaviour is consistent during the whole evolution.

\subsubsection{Dependence on time}\label{sec:dynamic}

\begin{figure}[b!]
    \centering
    \includegraphics[width=0.7\linewidth]{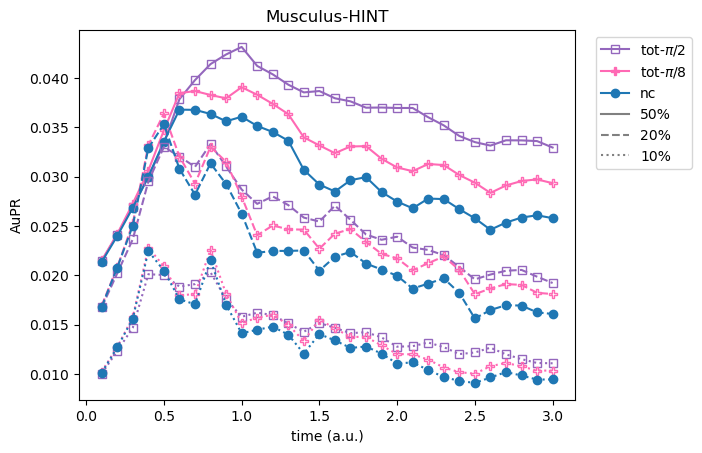}
    \caption{Evolution in time of the average AuPR on Musculus-HINT obtained through $k$-fold cross-validation for $10\%$, $20\%$ and $50\%$ link removal, as explained in the text. Labels follow the same nomenclature as Fig. \ref{fig:conv}; in addition, the suffix `$\pi/8$' indicates the case with phases sampled from the continuous interval $[-\pi/8,\pi/8]$.}
    \label{fig:tdep}
\end{figure}

As mentioned above, the problem of maximizing the AuPR requires an optimization on the hyperparameter $t$, which defines the evolution time of the walkers. {\color{black} From a physical perspective, relevant exploration times are expected to be of the order of the inverse spectral gap of the Hamiltonian, corresponding to the timescale needed for information to propagate beyond the immediate neighbourhood of the starting node. For the $k$-fold analysis we considered, the average spectral gaps at $10\%$, $20\%$ and $50\%$ link removal are $\Delta \lambda^{(10\%)}=(2.2\pm 0.3)$, $\Delta \lambda^{(20\%)}=(2.0\pm 0.5)$ and $\Delta \lambda^{(50\%)}=(1.6\pm 0.7)$. For shorter times the dynamics remains local, while for excessively long times the evolution becomes dominated by recurrences and high‑frequency interference, progressively washing out structural information.}

{\color{black}That said, in realistic PPI reconstruction tasks, an effective optimisation over $t$ cannot be performed, as the true set of missing links is unknown and cannot be used for validation. Consequently, rather than proposing a new strategy to optimise the evolution time $t$, here we analyse how sensitive the link‑prediction performance is to its choice, and whether chirality alleviates this limitation of quantum walk-based methods.} In Fig. \ref{fig:tdep} we show the dependency of the AuPR on time for Musculus-HINT at various link removal rates. For this analysis, we consider the second sampling method from Sec. \ref{sec:static} (suffix `$\pi/2$'), which we showed to be the more efficient in creating c-QW's with complementary evolutions. We also introduce a new method (suffix `$\pi/8$'), which samples the phases from a small continuous interval, i.e. $[-\pi/8,\pi/8]$. Similarly to the first method from Sec. \ref{sec:static}, the phases from the $\pi/8$-method can be any real number in the interval, but now they act as a perturbation on the non-chiral evolution.

Figure \ref{fig:tdep} shows that, for all percentages of link removal, the walkers require an initial time to leave the starting node and adequately explore the surrounding region. During this {\color{black}transient regime}, the AuPR increases rapidly, until it reaches a maximum at $t\lesssim 1$. At this point, we can imagine that the walkers have explored most of the topology of the graph. Afterwards, the AuPR starts decreasing {\color{black}due to the previously-mentioned high-frequency interference effects, which disrupt} the coherence of the quantum walk in its later stages. Eventually, the system reaches a saturation regime, with AuPR values that are significantly lower than those at the maximum.

At the optimal time and for low link removal rates ($10\%$ and $20\%$), the AuPR obtained with the $\pi/2$-sampling method is below those of the nc-QW or the $\pi/8$-method. The latter is instead capable of maintaining the non-chiral performance thanks to its perturbative nature. Instead, for high link removal rates ($50\%$) the $\pi/2$-method largely surpasses the others. Particularly, we observe an increase of $\sim 5.9\%$ for the $\pi/8$-method and $\sim 8.6\%$ for the $\pi/2$-method with respect to the maximum of the non-chiral curve. This proves that the best usage of the swarm of c-QW's is in particularly incomplete networks, such as PPI networks \cite{luck2020,dimitrakopoulos2022,kosoglu2024}. That said, at all link removal percentages either chiral methods improve the performance at later times ($t\gtrsim 1$). Consequently, in all cases the chiral swarm provides a more stable performance over time, and thus becomes the most viable option in real-life scenarios in which time optimization is not possible.  The $\pi/2$-method particularly so, due to the walkers' complementarity in evolution, mentioned previously in Sec. \ref{sec:static}.

{\color{black}Overall, while the optimal AuPR may occur at different times depending on the phase‑sampling strategy, the chiral swarms maintain high performance over a substantially broader temporal window, as evidenced by the larger integrated area under the AuPR-time curves.} Consequently, the choice over nc-QW's versus c-QW's, and over which sampling method to use, depends mainly on the intended use of the algorithm: specifically, on whether time  optimization is possible, or one-shot measurements at an arbitrary time are required. {\color{black}Specifically,} the $\pi/2$-method provides {\color{black}the most} stable performance over time, especially for high link removal rates. {\color{black}Conversely,} the $\pi/8$-method {\color{black}provides an optimal performance much more similar to the non-chiral case, and} seems to be a good trade-off between the two other options.


\subsection{Application on PPI networks}\label{sec:benchmark}
We then checked the performance of our algorithms on different and larger networks (see Table \ref{tab:param}), comparing our results with various other methods from literature. Among these, \textit{L3} was specifically designed to work on PPI networks, and is based on the weighted counting of paths of length three between nodes \cite{kovacs2019}. \textit{Preferential attachment} (PA) instead evaluates the score as the product of the degree of the nodes \cite{newman2001}. Then, \textit{common neighbour} (CN) analyses the number of neighbours shared by the nodes \cite{yao2016}. \textit{Adamic Adar} (AA) is similar to CN, but adds additional weight to less-connected neighbours \cite{adamic2003}. Finally, the \textit{structural perturbation method} (SPM) estimates the predictability on the adjacency matrix of the graph by adding perturbations to it \cite{lu2015}. For the latter we fixed $p^H=0.1$ and averaged results over 10 runs as in Ref. \cite{lu2015}.

\begin{figure}[h!]
\centering
    \includegraphics[width=0.8\linewidth]{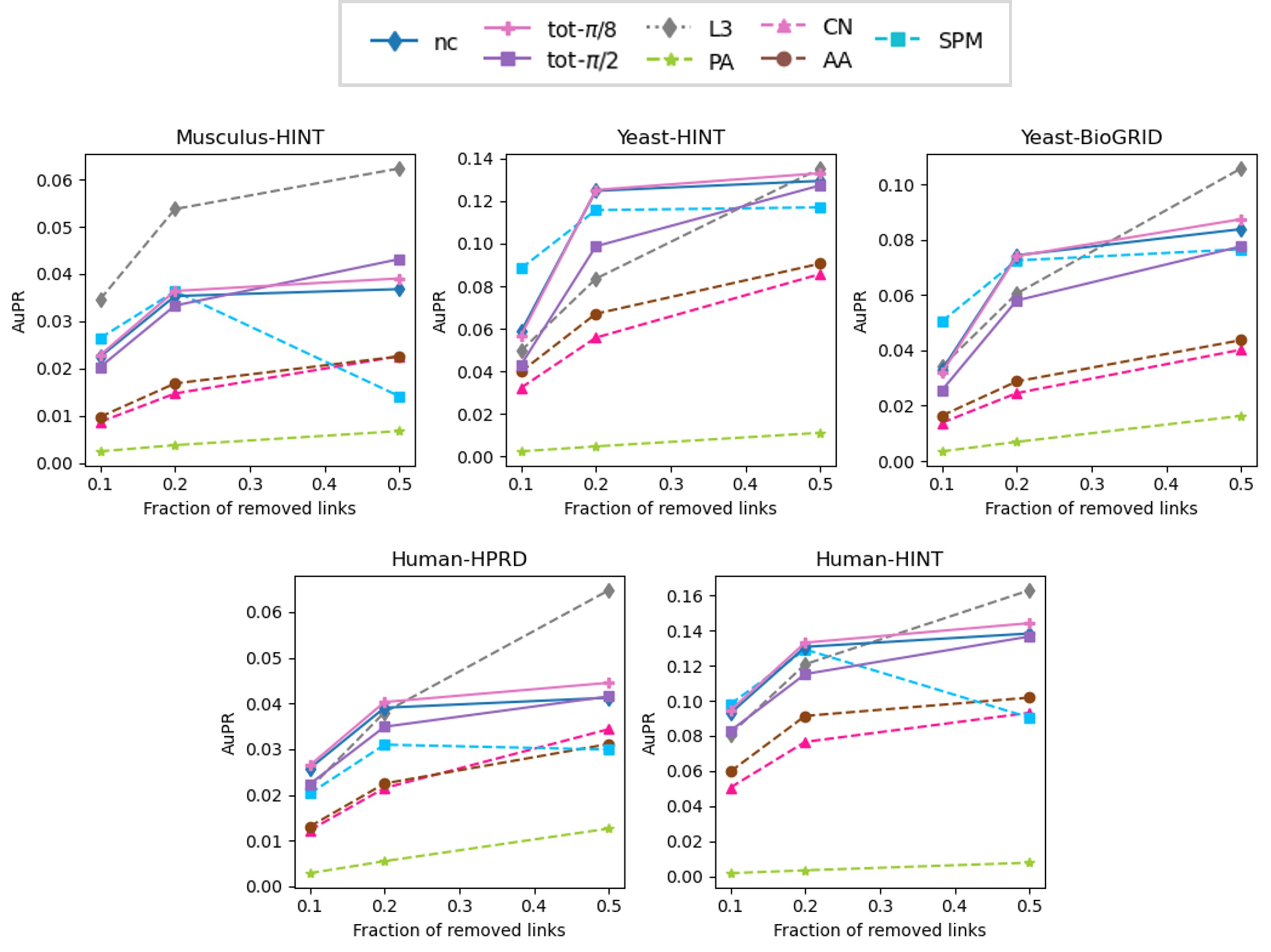}
    \caption{Average AuPR as a function of the fraction of removed links, obtained through repeated $k$-fold cross-validation, as explained in the text. All continuous lines refer to our link prediction algorithm using chiral and non-chiral QW's, while all dashed lines correspond to various other methods from literature. Labels follow the same nomenclature as previous Figures.}
    \label{fig:comp_benchmark_aupr}
\end{figure}

\begin{figure}[t]
\centering
    \includegraphics[width=0.8\linewidth]{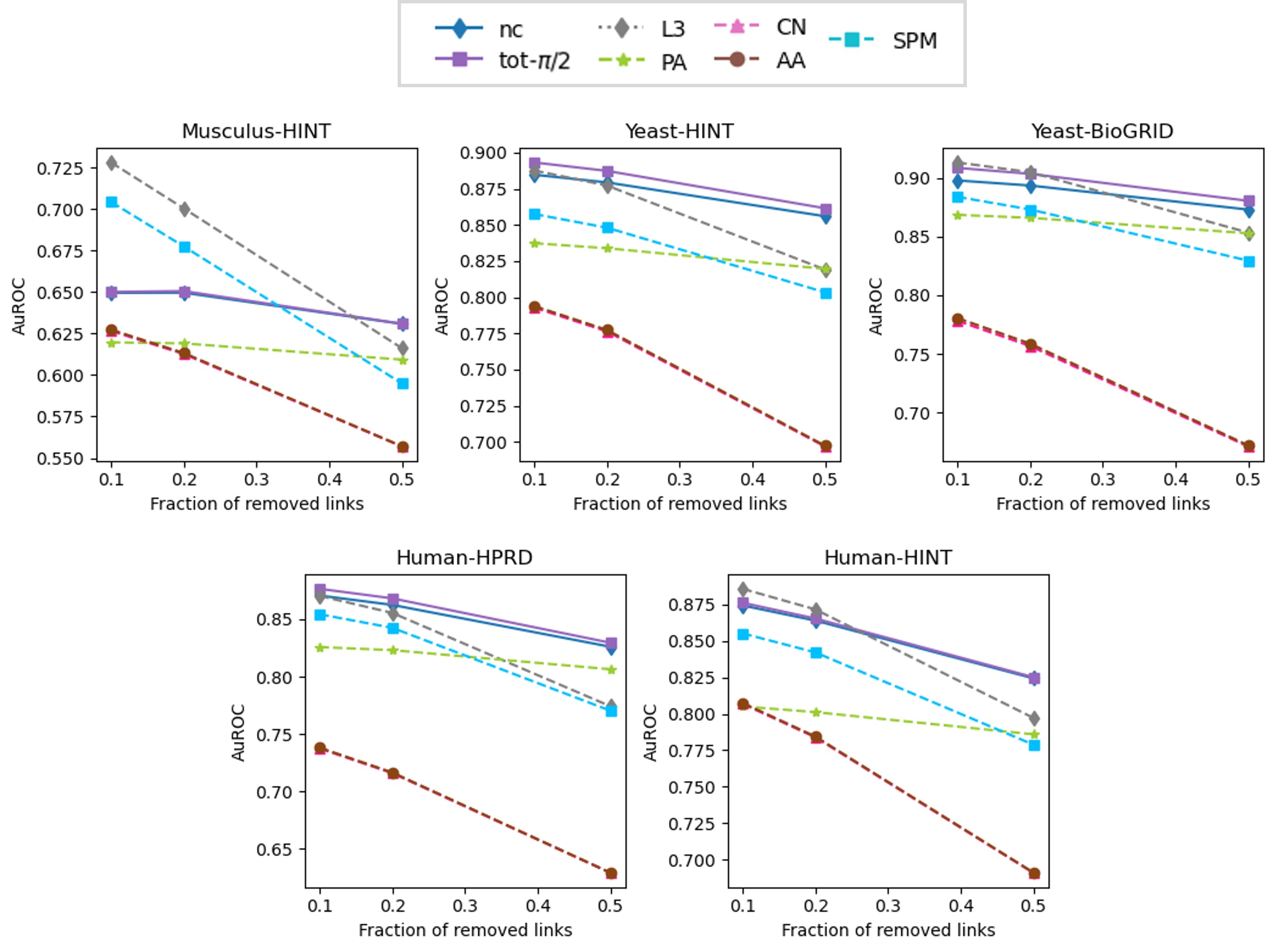}
    \caption{Average AuROC as a function of the fraction of removed links, obtained through repeated $k$-fold cross-validation, as explained in the text. All continuous lines refer to our link prediction algorithm using chiral and non-chiral QW's, while all dashed lines correspond to various other methods from literature. Labels follow the same nomenclature as previous Figures.}
    \label{fig:comp_benchmark_auroc}
\end{figure}

In Figs. \ref{fig:comp_benchmark_aupr} and \ref{fig:comp_benchmark_auroc} we report respectively the AuPR and AuROC for all cases. Notice that we only report on the AuPR at optimal times, to better understand in which cases the maximal performance can be improved with chirality. Otherwise, the same considerations as in Sec. \ref{sec:musculus} stand for all other networks as well, and the $\pi/2$-method is the one that provides the most stable performance, and the higher AuPR at the reference time of $t=1$ (not shown). As for the AuROC, its time dependence is negligible and does not affect the ordering of the performances between the QW-methods, so that we directly plot the results at $t=1$.

The curves of the AuPR in Fig. \ref{fig:comp_benchmark_aupr} show good performance for the QW-methods with respect to the other most common algorithms from literature. Specifically, the only two other methods that prove to be always competitive with them are SPM and L3. Both of these are particularly efficient in PPI networks. Indeed, the perturbative nature of the former is effective in predicting missing parts of networks \cite{lu2015}, while the latter  was specifically designed to capture the basic principles that drive protein interactions \cite{kovacs2019}. That said, in all cases the QW-algorithms are able to match or even outperform at least one between L3 and SPM, if not both (e.g. Yeast-HINT at 20\% link removal).

That said, for all the studied networks, the previous considerations from Sec. \ref{sec:dynamic} on the behavior of the QW-methods' optimal AuPR are confirmed apart for one exception. That is, at 50\% link removal, the $\pi/2$-method only outperforms the others in Musculus-HINT, while in all other networks it is hardly able to match the non-chiral performance. Notice that the Musculus network is the denser one and with the lowest clustering coefficient (see Table \ref{tab:param}). Consequently, this suggests that chirality is less efficient on those networks that, instead, are globally sparser but locally denser, as is the case for the two Yeast networks and the two Human ones. 

On the other hand, the $\pi/8$-method always shows a slight but definite improval at high link removal rates with respect to the non-chiral method, while still matching it at lower rates. This is because the $\pi/8$-method is, as mentioned previously, a perturbation on the non-chiral method. Consequently, though still maintaining the same flexibility of a chiral algorithm, each c-QW has an evolution that is more similar to the non-chiral one. 
Specifically, the $\pi/8$-method improves the AuPR by $2.8\%$ and $4.3\%$ respectively for Yeast-HINT and Yeast-BioGRID at 50\% link removal; and by $8.0\%$ and $4.3\%$ respectively for Human-HPRD and Human-HINT at 50\% link removal.

As for the AuROC, Fig. \ref{fig:comp_benchmark_auroc} shows that, in almost all cases, the QW-algorithms match/outperform the other best method from literature, i.e. L3. The only striking exception is the Musculus-HINT network at low link removal rate. However, even in this network, the QW-method shows the overall best performance at $50\%$ link removal. This behavior confirms that, as mentioned previously in Sec. \ref{fig:tdep}, the QW-method with either the nc-QW or the swarm of c-QW's is best used when the percentage of missing links is high, as is the case for current PPI databases in general \cite{luck2020,dimitrakopoulos2022,kosoglu2024}. Additionally, the chiral $\pi/2$-method always increases, though often slightly, the non-chiral performance. The $\pi/8$-method (not shown for clarity of the picture) is instead always in between the non-chiral performance and the chiral one with the $\pi/2$-method, thus confirming its nature as a middle ground between the two other methods. The most relevant enhancements are observed in the Yeast-HINT network ($0.96\%$ at $10\%$ link removal rate) and in the Yeast-BioGRID network ($1.19\%$ at $10\%$ link removal rate).

\subsubsection{Version comparison}\label{sec:version}

Up to now, we have evaluated the link prediction performance through $k$-fold cross-validation on specific versions of network databases. That said, an additional benchmark consists of comparing different versions of the same databases, by applying link prediction to earlier versions of the networks studied in this paper. In order to do this, we considered the versions of the Yeast-HINT, Yeast-BioGRID and Human-HINT databases previously studied in Ref. \cite{goldsmith2023}. This strategy has low statistical significance, since it relies on a single way in which the databases historically evolved. However, it provides a more realistic assessment of the performance of our algorithms on real-world scenarios, and helps identify which algorithms are more robust to structural changes over time. 

\begin{table}[t]
    \centering
    \begin{tabular}{|r|c|c|c|}
    \hline
        & Yeast-HINT & Yeast-BioGRID & Human-HINT \\
    \hline
    CN     &  0.0138 & -- & \textbf{0.279} \\
    \hline
    SPM     & 0.0352 & 0.0186 & 0.233 \\
    \hline
    AA     & 0.0112 & -- & 0.276 \\
    \hline
    L3     & 0.0125 & -- & 0.251\\
    \hline
    PA     & -- & -- & 0.0791\\
    \hline
    Non-chiral     & 0.0706 & 0.0170 & 0.219\\
    \hline
    Chiral-$\pi/2$     & 0.0250 & \textbf{0.0285} & 0.120\\
    \hline
    Chiral-$\pi/8$     & \textbf{0.0726} & 0.0120 & 0.212\\
    \hline
    \end{tabular}
    \caption{AP@$k$ obtained through the comparison of different versions of the Yeast-HINT, Yeast-BioGRID and Human-HINT databases, for $k=100$. The times of the QW evolutions were fixed to values close to the optimal times considered for Fig. \ref{fig:comp_benchmark_aupr}, i.e. $t=0.3$ for Yeast-HINT; $t=0.2$ for Yeast-BioGRID; and $t=0.4$ for Human-HINT. No performance is reported in those cases where no links were found in the first $k$ predictions; the best performance for each column is highlighted in bold. The suffixes of the chiral methods follow the nomenclature used throughout the paper.}
    \label{tab:apk}
\end{table}

Here we evaluate the performance on version comparison through the average precision at $k$ (AP@$k$), computed as:
\begin{equation}
    {\rm AP@}k = \frac{1}{r_k}\sum_{i=1}^k \frac{r_i}{i} \cdot {\rm rel}(i)\, ,
\end{equation}
where $r_k$ ($r_i$) is the number of relevant items, i.e. of correctly predicted links, in the top $k$ ($i$) predictions; the ratio $r_k/k$ is also called precision at $k$ (P@$k$); and rel$(i)$ is an indicator function equal to 1 if the item of rank $i$ is relevant, and 0 otherwise. The AP@$k$ measures how well a model ranks the true future links among its top $k$ predictions, reflecting both the accuracy and the relevance of the predicted links. This is especially useful in temporal evaluations, where the goal is to predict the links that actually formed in a newer version of the network. Additionally, by focusing on the top $k$ predictions, the AP@$k$ aligns with real-world scenarios where only a limited number of high-confidence predictions are possible to test, as is the case in PPI network reconstruction.

Table \ref{tab:apk} reports the AP@$k$ for the three networks under consideration and for all the algorithms previously introduced, considering $k=100$. In both Yeast databases, the best performance corresponds to one of the two chiral methods. Specifically, the increase with respect to the non-chiral one is relatively small for Yeast-HINT with the $\pi/8$-method ($\sim 2.8\%$), but it is especially high on Yeast-BioGRID with the $\pi/2$-method ($\sim 67 \%$). In these two networks, all QW-based algorithms are particularly better than L3, which consistently outperformed all other methods at high link removal rate in the results from Fig. \ref{fig:comp_benchmark_aupr}. That is to say that, although L3 performs the best on cross-validation on a single version of a database, it does not necessarily imply the best performance in a real-life scenario. That said, the results from the Human-HINT database show that, in this network, the performance from all methods is much higher than in the Yeast networks. Especially CN, which instead was performing very poorly in Figs. \ref{fig:comp_benchmark_aupr} and \ref{fig:comp_benchmark_auroc}. Here, though the QW performance is slightly lower, it is still in the same order of magnitude as the others.

Overall, considering the values of the AP@$k$ in absolute terms, chirality does not show a substantial advantage in scenarios where other methods already achieve high performance. However, it proves to be particularly valuable in enhancing the non-chiral method in cases where most other approaches fall short, highlighting its role as a complementary asset within the broader landscape of link prediction techniques.


\section{Conclusions}\label{outro}
In this paper we have generalized a quantum-based approach to link prediction in complex networks, through the use of swarms of chiral quantum walks. Thanks to randomly-sampled complex phases in the Hamiltonian generator, each walker of the swarm possesses a different directional bias that complements the others in the exploration of the networks. While a single non-chiral quantum walk typically outperforms an individual chiral walk due to its uniform exploration, a well-constructed swarm of chiral walkers can explore the network more thoroughly and stably over time. This advantage becomes particularly relevant in real-world scenarios where the optimal evolution time is unknown. {\color{black}Indeed, in practical applications, this swarm-based strategy reduces the sensitivity of the prediction performance to the precise choice of $t$, enabling near-optimal results even when the evolution time cannot be finely tuned.} By employing targeted phase sampling strategies—such as maximizing dynamical diversity—chiral swarms achieve greater robustness, sometimes outperforming the non-chiral case at optimal time. When the latter case does not occur, small chiral perturbations offer a practical compromise, preserving the strengths of the non-chiral evolution while enhancing stability over the evolution time.


We then tested our methods across multiple real-world PPI networks, showing that the QW-based algorithms consistently matched or outperformed classical algorithms such as L3 and SPM, particularly in correspondence of high link removal rates. Maximizing the distance between each walker proved most effective in sparse and low-clustering networks. Instead, small perturbations of the non-chiral evolution offered a reliable compromise in more locally dense topologies, notably improving the AuPR by up to 8.0\% in the Human-HPRD database. Furthermore, in version comparison experiments—where predictions were made on future links from older network snapshots—chiral methods showed superior performance especially when most other methods failed at retrieving any link in the top 100. These results highlight the complementary role of chirality in enhancing link prediction, particularly in dynamic and largely incomplete network scenarios.

\section*{Acknowledgments}
This work has been partially supported by EU (Next Generation Europe) and MUR through the project QWEST-CUPG53D23006270001.

\appendix

\section{Convergence of AuROC}\label{sec:auroc}

\begin{figure}[ht]
    \centering
    \includegraphics[width=0.5\linewidth]{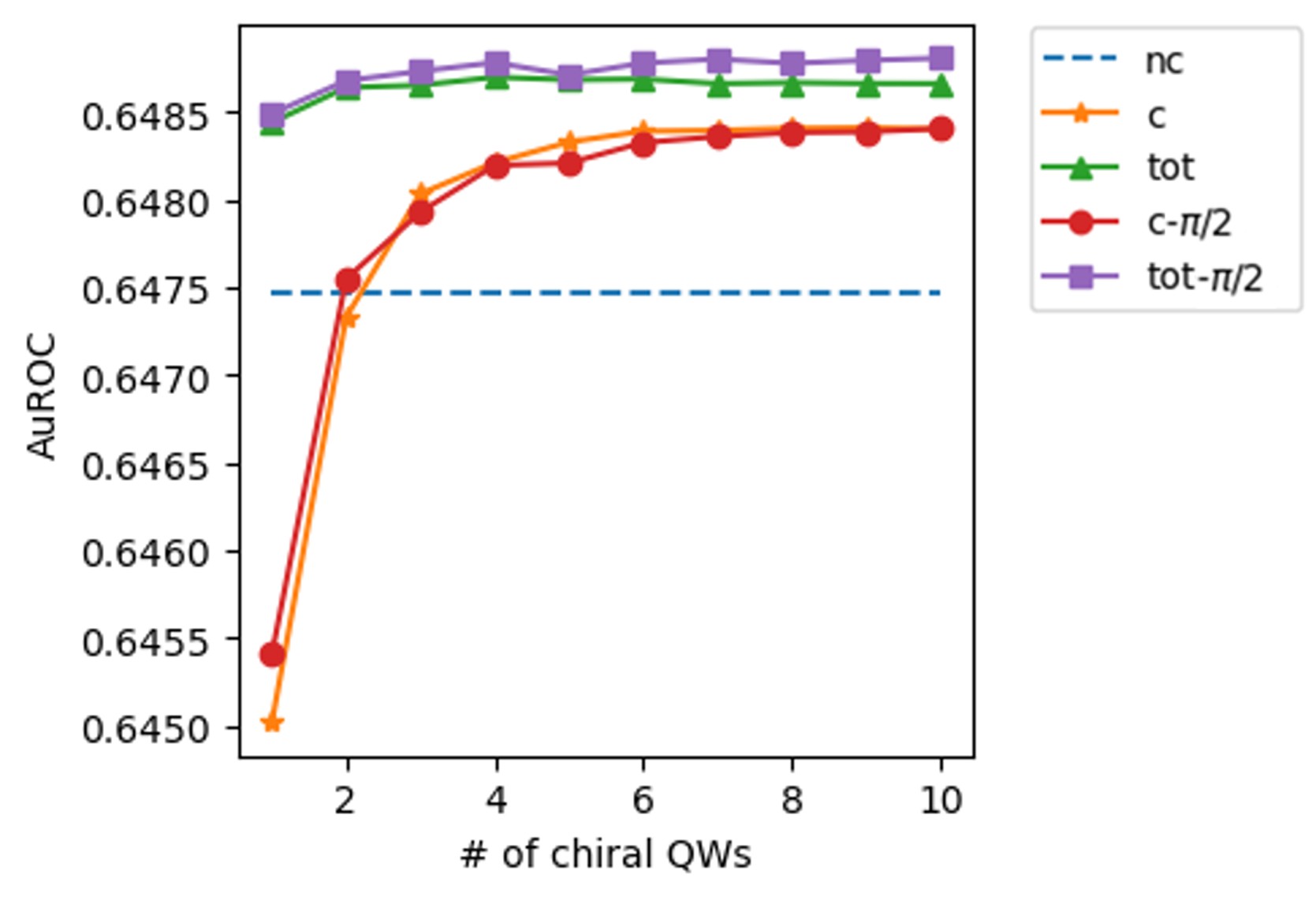}
    \caption{Convergence of the average AuROC at $10\%$ link removal obtained from 10-fold cross-validation on the Musculus-HINT network. Labels follow the same nomenclature as Fig. \ref{fig:conv}.}
    \label{fig:conv_auroc}
\end{figure}

Figure \ref{fig:conv_auroc} shows the AuROC at $10\%$ link removal at unitary time from the 10-fold cross-validation on the Musculus-HINT network, as a function of the number of chiral walks considered. The curves present the same convergent behavior as the AuPR, reported in Fig. \ref{fig:conv}(b), and the same considerations apply, although the numerical variation among curves is much smaller. 

\section{Dynamical distance between walker evolutions}\label{sec:distance}

\begin{figure}[ht]
    \centering
    \includegraphics[width=0.9\linewidth]{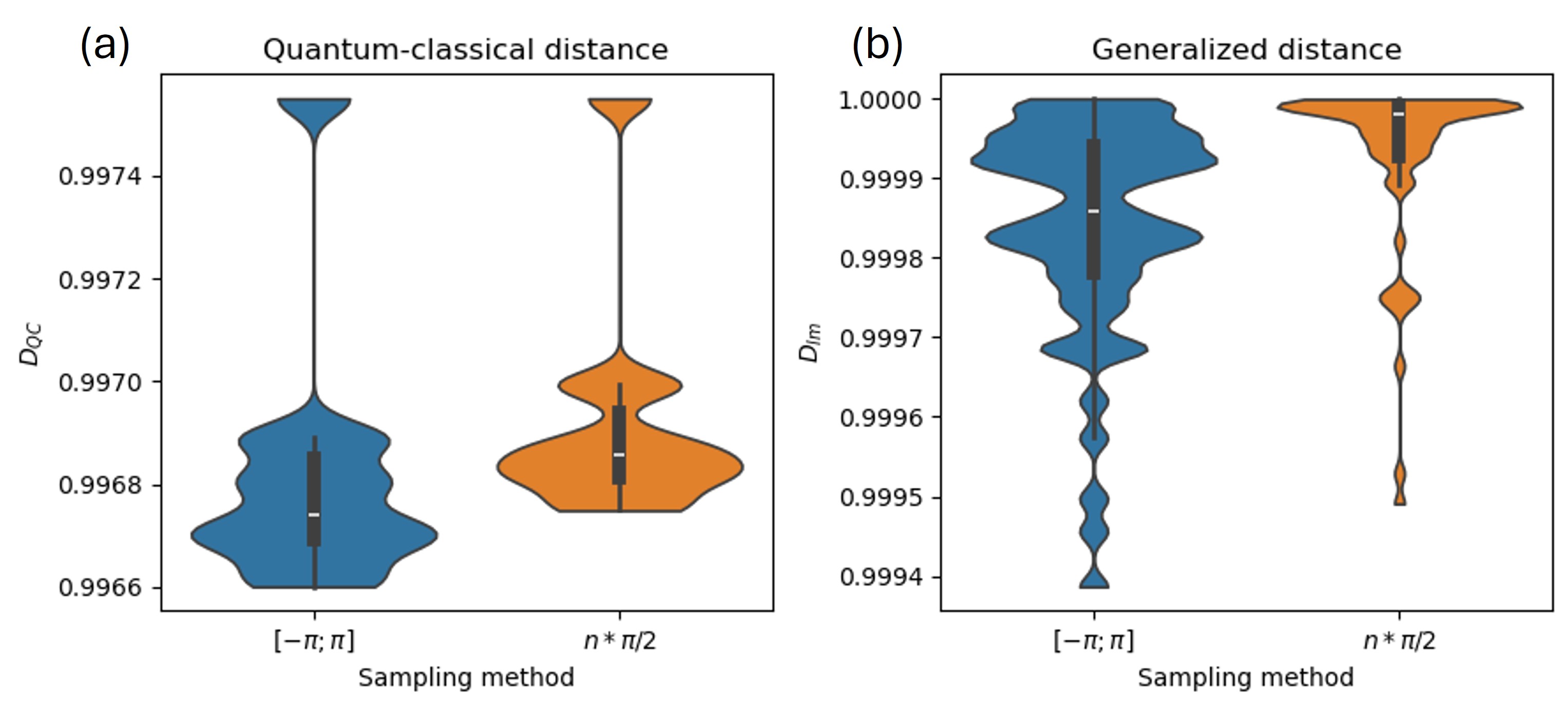}
    \caption{Violin plots showing the distribution of (a) the quantum-classical distance $\mathcal{D}_{QC}$ between a QW and the corresponding RW and (b) the generalized distance $\mathcal{D}_{lm}$ between pairs $(l,m)$ of QW's, both as described in the text and maximized on all possible starting nodes $\left|j\right\rangle$. Two swarms built with different sampling methods were considered, both made up of $10$ c-QW's and the corresponding nc-QW. The first sampling methods (blue) considered phases in the continuous interval $[-\pi,\pi]$; the second (orange) considered only multiples of $\pi/2$.}
    \label{fig:dist}
\end{figure}

 A possible tool to determine the difference in evolution between a QW and the corresponding RW both starting from the same localized state $\left|j\right\rangle$ is the so-called quantum-classical distance \cite{frigerio2022}, defined at time $t$ as
\begin{align}\label{eq:qc-distance}
    \mathcal{D}^j_{QC}(t)=1-\mathcal{F}[\hat{\sigma}_{j}(t),\hat{\rho}_j(t)] = 1-\sum_{k=1}^n P^{\rm RW}_{jk}(t)P^{\rm QW}_{jk}(t)\, ,
\end{align}
where $\hat{\sigma}_j$ and $\hat{\rho}_j$ are the classical and quantum density matrices at time $t$, respectively, and $\mathcal{F}[\hat{\sigma}_{j},\hat{\rho}_j]={\rm Tr}[(\sqrt{\hat{\sigma}_j}\hat{\rho}_j\sqrt{\hat{\sigma}_j})^{1/2}]^2$ is the fidelity between states. In Eq. \eqref{eq:qc-distance} we used the fact that $\hat{\rho_j}$ is a pure state. A global distance, independent of the initial state $\left|j\right\rangle$, can be defined as $\mathcal{D}_{QC}(t)=\max\{\mathcal{D}^j_{QC}(t)\}$. In all the cases we considered in this paper, this distance, when evaluated between each walker and the corresponding RW, is extremely high ($\mathcal{D}_{QC}>0.99$). Additionally, we show in Fig. \ref{fig:dist}(a) the distributions of this distance for two swarms evolving with $t=1$ on the Musculus-HINT network at $10\%$ link removal. Each swarm was built with one of the two sampling methods described in Sec. \ref{sec:static}, that is, either with random phases in the whole continuous interval $[-\pi,\pi]$ or with multiples of $\pi/2$. As we see, the only case which distinguishes itself in both distributions is the nc-QW, which does not possess any directional bias and thus behaves slightly differently with respect to the other c-QW's. That said, the two distributions are mostly overlapping, showing very similar averages and standard deviation. This suggests that this quantity alone does not provide us with any useful insights into the different behaviors of different sampling methods.

Consequently, we generalize Eq. \eqref{eq:qc-distance} to the definition of a distance between the evolution of any two QW's, either chiral or non-chiral, evolving on the same network. Considering, as before, the case in which the two walkers $l$ and $m$ start from the same localized state $\left| j\right\rangle$:
\begin{align}\label{eq:distance}
    \mathcal{D}^j_{lm}(t)=1-\mathcal{F}[\hat{\rho}_{j,l}(t),\hat{\rho}_{j,m}(t)] = 1-\left|\left\langle\psi_{j,l}(t)|\psi_{j,m}(t)\right\rangle\right|^2 = 1- |\langle j|\, \mathcal{U}_l^\dagger(t) \, \mathcal{U}_m(t)|j\rangle|^2\, ,
\end{align}
where we used the fact that both states are pure. Its global equivalent on the set of all possible localized initial states is $\mathcal{D}_{lm}(t)=\max\{\mathcal{D}^j_{lm}(t)\}$.

Figure \ref{fig:dist}(b) shows the distributions of the quantity $\mathcal{D}_{lm}(t=1)$ for the same cases as in Fig. \ref{fig:dist}(a). In this case, there is no appreciable difference in either considering the distance between a c-QW and the nc-QW, or between two c-QW's. However, now the two distributions are noticeably different, with the $\pi/2$-method showing a much more squeezed distribution around 1, which is the maximum possible value of the distance. This proves that, while still showing similar distances with respect to the RW evolution, the $\pi/2$ method produces a set of evolutions that are as different as possible from one another, thus exploring the possible configuration space in a much more efficient manner with respect to the $[-\pi,\pi]$-method.

\bibliographystyle{unsrt}
\bibliography{bibliography.bib}

\end{document}